\documentclass[
 reprint,
superscriptaddress,
 amsmath,amssymb,
 aps,
 prb,
]{revtex4-2}

\usepackage{graphicx}
\usepackage{float}
\usepackage{dcolumn}
\usepackage{bm}
\usepackage{bbm}
\usepackage{braket}
\usepackage{xcolor}
\usepackage{physics}
\usepackage{hyperref}
\usepackage{leftindex}

\makeatletter
\DeclareFontFamily{OMX}{MnSymbolE}{}
\DeclareSymbolFont{MnLargeSymbols}{OMX}{MnSymbolE}{m}{n}
\SetSymbolFont{MnLargeSymbols}{bold}{OMX}{MnSymbolE}{b}{n}
\DeclareFontShape{OMX}{MnSymbolE}{m}{n}{
    <-6>  MnSymbolE5
   <6-7>  MnSymbolE6
   <7-8>  MnSymbolE7
   <8-9>  MnSymbolE8
   <9-10> MnSymbolE9
  <10-12> MnSymbolE10
  <12->   MnSymbolE12
}{}
\DeclareFontShape{OMX}{MnSymbolE}{b}{n}{
    <-6>  MnSymbolE-Bold5
   <6-7>  MnSymbolE-Bold6
   <7-8>  MnSymbolE-Bold7
   <8-9>  MnSymbolE-Bold8
   <9-10> MnSymbolE-Bold9
  <10-12> MnSymbolE-Bold10
  <12->   MnSymbolE-Bold12
}{}

\let\llangle\@undefined
\let\rrangle\@undefined
\DeclareMathDelimiter{\llangle}{\mathopen}%
                     {MnLargeSymbols}{'164}{MnLargeSymbols}{'164}
\DeclareMathDelimiter{\rrangle}{\mathclose}%
                     {MnLargeSymbols}{'171}{MnLargeSymbols}{'171}
\makeatother

\begin{document}

\title{Heisenberg-limited metrology from the quantum-quench dynamics of an anisotropic ferromagnet }

\author{Z.~M.~McIntyre}
\email{zoe.mcintyre@unibas.ch}
\affiliation{Department of Physics, University of Basel, Klingelbergstrasse 82, 4056 Basel, Switzerland}
\author{Ji Zou}%
\email{ji.zou@unibas.ch}
\affiliation{Physics Department, King Fahd University of Petroleum and Minerals, 31261, Dhahran, Saudi Arabia}
\affiliation{Quantum Center, KFUPM, Dhahran, Saudi Arabia}
\author{Jelena Klinovaja} 
\affiliation{Department of Physics, University of Basel, Klingelbergstrasse 82, 4056 Basel, Switzerland}
\author{Daniel Loss} 
\affiliation{Department of Physics, University of Basel, Klingelbergstrasse 82, 4056 Basel, Switzerland}
\affiliation{Physics Department, King Fahd University of Petroleum and Minerals, 31261, Dhahran, Saudi Arabia}
\affiliation{Quantum Center, KFUPM, Dhahran, Saudi Arabia}
\affiliation{RDIA Chair in Quantum Computing}

\date{\today}

\begin{abstract}
    The emerging field of quantum magnonics seeks to understand and harness the quantum properties of magnons---quantized collective spin excitations in magnets. Squeezed magnon states arise naturally as the equilibrium ground states of anisotropic ferromagnets and antiferromagnets, representing an important class of nonclassical magnon states. In this work, we show how a qubit-conditioned quantum quench of an anisotropic ferromagnet can be used for Heisenberg-limited parameter estimation based on measurements of the qubit only. In the presence of ground-state squeezing, the protocol yields information about the eigenmode frequency of the coupled magnon-qubit system, whereas no information is gained in the absence of such squeezing.  The protocol therefore leverages genuine quantum correlations in the form of magnonic squeezing while simultaneously relying on the equilibrium character of this squeezing---a feature distinctive to magnetic systems.
\end{abstract}

\maketitle

\section{Introduction} 

The quantum nature of magnons, the quanta of collective spin excitations in magnets, has attracted intense interest in recent years, giving rise to the emerging and rapidly developing field of \textit{quantum magnonics}, see Refs.~\cite{yuan2022quantum,chumak2022advances} for reviews. Magnons hold promise as information carriers in quantum networks~\cite{wang2024nanoscale}, and significant progress has been made, not only theoretically but experimentally as well, towards the coherent control and measurement of quantum magnonic states via qubits~\cite{tabuchi2015coherent,lachance2017resolving,lachance2020entanglement,xu2023quantum,rani2025high,weng2026magnon}.

The maximum precision that can be achieved in estimating an unknown parameter $\phi$ using a classical coherent state $\ket{\alpha}$ is given by the standard quantum limit (SQL)~\cite{giovannetti2006quantum}, $\delta\phi=1/\sqrt{NM}$, with $M$ giving the number of independent measurements and $N=\vert \alpha\vert^2$ giving the average number of particles in the state. The fundamental goal of quantum metrology is to estimate $\phi$ from measurements of a quantum state with a precision scaling like $N^{-\nu}$ with $\nu>1/2$, outperforming the SQL. The benefits of using squeezed vacuum states for quantum metrology have been widely studied in the context of interferometry with quantum states of light~\cite{caves1981quantum,lang2013optimal,pezze2008mach,nielsen2023deterministic}, with Heisenberg-limited scaling $\delta\phi=1/(N\sqrt{M})$ having been demonstrated experimentally~\cite{nielsen2023deterministic}. Most famously, squeezed vacuum has been used by LIGO to enhance the sensitivity of gravitational wave detection in a Michelson interferometry setup~\cite{tse2019quantum}. Squeezed spin states of atomic ensembles~\cite{kitagawa1993squeezed} can also be used for quantum-enhanced parameter estimation in atomic Ramsey interferometry~\cite{wineland1994squeezed,gross2012spin,pezze2018quantum}.

\begin{figure}
    \centering
    \includegraphics[width=\linewidth]{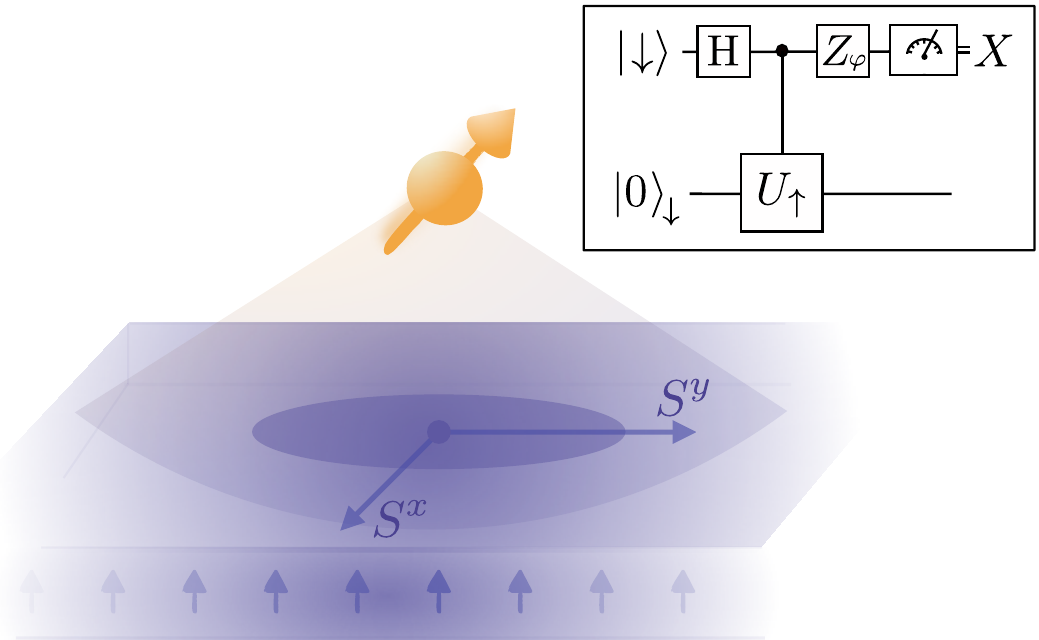}
    \caption{Schematic of a spin qubit coupled to the Kittel mode of an anisotropic ferromagnet. The ground state of the Kittel mode is a squeezed state. Inset: Protocol expressed as a circuit. The magnon mode is allowed to equilibrate while in contact with the qubit held in $\ket{\downarrow}$. The squeezed magnons used for parameter estimation are produced by subjecting the magnon system to a quantum quench conditioned on the state of the qubit being $\ket{\uparrow}$.} 
    \label{fig:sketch}
\end{figure}

Squeezed light and spin-squeezed ensembles are typically generated dynamically using strategies like Feshbach-resonance-induced interactions~\cite{gross2010nonlinear}, cavity-mediated interactions~\cite{sorensen2002entangling}, and parametric down-conversion~\cite{wu1986generation}. In stark contrast, squeezed magnonic states exist ubiquitously in nature, being found for instance in anisotropic ferromagnets~\cite{kamra2016super,kamra2016magnon,zou2020tuning} and antiferromagnets~\cite{kamra2019antiferromagnetic}. In these systems, squeezing emerges as an intrinsic feature of the magnetic ground state proceeding from energy minimization~\cite{kamra2020magnon}. Put differently, squeezed states of photonic or atomic systems are typically \textit{excited} states, which must be actively generated and then sustained against loss, whereas in magnetic systems, they are nonclassical resources that may already present at equilibrium. Despite growing theoretical understanding and recent proposals for detecting such states~\cite{romling2023resolving,romling2024quantum}, this equilibrium origin makes it challenging to access and harness magnon squeezing for useful applications. 

A key impediment to leveraging magnon squeezing for quantum metrology is the development of concrete measurement strategies enabling sub-SQL parameter estimation. For interferometry with photonic states, number-resolving measurements~\cite{pezze2008mach,lang2013optimal,eaton2023resolution}, parity measurements~\cite{anisimov2010quantum,seshadreesan2011parity}, and homodyne detection~\cite{gard2017nearly,mcintyre2024homodyne} can all be implemented using existing technologies. For atomic interferometers, the required atomic population measurements have been realized~\cite{pezze2018quantum}, and the atomic analog of homodyne detection has been demonstrated experimentally as well~\cite{gross2011atomic}. A cavity-magnonic metrology scheme was presented and analyzed theoretically in  Ref.~\cite{wan2024quantum}; however, entanglement between the cavity and magnon mode generated for the purpose of measuring the magnon via the cavity may in certain situations adversely affect the usefulness of magnon squeezing in achieving sub-SQL precision.

In this work, we show how a qubit-conditioned quantum quench of the magnon mode could be used for Heisenberg-limited estimation of an unknown frequency based on measurements of the qubit. Such measurements have been realized for qubits coupled to magnons~\cite{tabuchi2015coherent,lachance2017resolving,lachance2020entanglement,xu2023quantum}. Conditioned on the qubit state, the quench will create real, squeezed magnons out of the Bogoliubov vacuum in a process similar to the dynamical Casimir effect~\cite{dodonov2020fifty}. Under free evolution, the interaction of the qubit with these squeezed magnons will lead to super-resolution~\cite{resch2007time} of the unknown parameter in the qubit coherence, enabling Heisenberg-limited parameter estimation via X-basis qubit readout. Crucially, no information about the unknown parameter is obtained in the absence of ground-state (equilibrium) squeezing of the magnon mode. This protocol therefore harnesses not only quantum effects in the form of squeezing, but also the equilibrium nature of this squeezing.

\section{Model} 

We consider an anisotropic ferromagnet exchange-coupled to a spin qubit (Fig.~\ref{fig:sketch}). The microscopic Hamiltonian $H_{\mathrm{sys}}$ describing this system (setting $\hbar=1$) is 
\begin{align}
\begin{aligned}
    H_{\mathrm{sys}}&=\frac{\Delta}{2} \sigma_z-J\sum_{\langle i, j\rangle} {\vb S}_i\cdot {\vb S}_j -K_z \sum_i(S_i^z)^2- K_y \sum_i(S_i^y)^2\\
    &-\lvert\gamma\rvert \mu_0 h  \sum_i{S}_i^z-J_{\mathrm{int}}\sum_i \vb S_i\cdot {\vb*\sigma}.
\end{aligned}
\end{align}
Here, $\Delta$ and $\sigma_z=\ketbra{\uparrow}-\ketbra{\downarrow}$ are the spin-qubit splitting and Pauli-Z operator, $J>0$ denotes the ferromagnetic exchange coupling, and $K_z, K_y>0$ represent easy-axis anisotropies along the $z$- and $y$-directions, respectively. The parameter $h$ denotes an external magnetic field applied parallel to the $z$-axis ($\bm{h}=h\hat{z}$), $\gamma=-28\:$GHz/T is a gyromagnetic ratio, $\mu_0$ is the vacuum magnetic permeability, and $J_{\mathrm{int}}$ is the interfacial exchange coupling strength between the spin qubit and ferromagnet. We focus on the Kittel ($\bm{k}=0$) mode of the magnet, described by the magnon creation operator $m^\dagger$, and we work in the detuned regime where the qubit is far off-resonant from the magnon frequency, allowing flip-flop terms in the exchange interaction to be neglected.  The effective Hamiltonian for the qubit and Kittel mode is then given by~\cite{skogvoll2021tunable,romling2023resolving}
\begin{align}
&H=H_0+H_{\mathrm{int}},\\
    &H_0=  \frac{\Delta}{2}\sigma_z+\omega_0 m^\dagger m+\Omega( m^2 + \mathrm{h.c.}),\label{free-hamiltonian}\\
    &H_{\mathrm{int}}=\chi m^\dagger m\sigma_z.\label{dispersive-coupling}
\end{align}
Here $\omega_0=2SK_z-SK_y+\lvert\gamma\rvert \mu_0 h$ is the magnon-mode frequency, which is tunable via the external field $h$, and $\Omega=SK_y/2$ with $S$ denoting the length of the spin. The anisotropy $K_y$ that generates $\Omega$ need not be magnetocrystalline: It can be supplied by the shape anisotropy of a magnet whose two axes transverse to $\bm{h}$ are inequivalent, for which $\Omega=\lvert\gamma\rvert\mu_0 M_s (N_x-N_y)/4$ with $N_i$ the demagnetizing factors and $M_s$ the saturation magnetization. For yttrium iron garnet, $\lvert\gamma\rvert\mu_0 M_s/2\pi\approx 4.9$~GHz. The value $\Omega/2\pi=0.5$~GHz used below would then correspond to $N_x-N_y\approx 0.4$, i.e.~to a moderately flattened ellipsoid. We emphasize that a sphere ($N_x=N_y=N_z$) or spheroid of revolution about $\bm{h}$ ($N_x=N_y$) both give $\Omega=0$. To obtain squeezing from shape anisotropy, an ellipsoid with three distinct semi-axes is therefore required. Such a geometry is in fact better suited to the present treatment than a thin film since the internal demagnetizing field of an ellipsoid is uniform. This in turn justifies the description of the magnet by the single Kittel mode.

The dispersive coupling $\chi$ in Eq.~\eqref{dispersive-coupling} originates from the Ising term in the exchange interaction and is given by $\chi=J_{\mathrm{int}}N_{\mathrm{int}}\vert \psi\vert^2/(2 N)$, where $N_{\mathrm{int}}$ is the number of interfacial spins, $N$ is the total number of sites in the ferromagnet, and $\vert\psi\vert^2$ is the electronic wavefunction of the qubit averaged over the interface~\cite{skogvoll2021tunable, romling2023resolving}. 
The aforementioned flip-flop terms can be neglected when $g/|\Delta-\omega_0|\ll 1$, where $g$ is the strength of the flip-flop terms. Alternately, a Schrieffer--Wolff elimination of the flip-flop interaction only produces an additional dispersive shift $\sim g^2/(\Delta-\omega_0)m^\dagger m\sigma_z$, which can be absorbed into $\chi$ without changing the structure of the Hamiltonian.
Although Eqs.~\eqref{free-hamiltonian}-\eqref{dispersive-coupling} were derived starting from a microscopic model $H_{\mathrm{sys}}$ in which the coupling between the (spin) qubit and the ferromagnetic spins originates from the exchange interaction, a dispersive coupling $m^\dagger m\sigma_z$ can also be realized for superconducting qubits by coupling both the superconducting qubit and Kittel mode to a microwave cavity mode~\cite{lachance2017resolving,lachance2020entanglement,rani2025high,weng2026magnon}. In Appendix \ref{app:NV}, we also show how a dispersive coupling could be obtained for an NV-center spin interacting with the stray field of a nearby nanomagnet. 

The $\mathrm{U}(1)$-symmetry-breaking term $\Omega(m^2+\mathrm{h.c.})$ violates magnon-number conservation and changes the nature of the magnetic ground state: Instead of the vacuum Fock state $\ket{0}$ (defined by $m \ket{0} = 0$), the ground state of $H$ is given by a nonclassical superposition of even-parity magnon states known as a squeezed magnon vacuum. To be exact, the ground state $\ket*{0}_{\!\sigma}$ ($\sigma=\:\uparrow,\downarrow$) of $H$ depends on the state $\ket{\sigma}$ of the qubit and is given by $\ket*{0}_{\!\sigma}=S(r_{\sigma})\ket{0}$, where $S(z)=\exp{(z^*m^2-\mathrm{h.c.})/2}$ is a squeezing operator acting on the magnon mode $m$. The strength $r_{\sigma}$ of the squeezing is qubit-state-dependent and given by $r_{\sigma}=(1/2)\mathrm{arctanh}[2 \Omega/\omega_{\mathrm{eff},\sigma}]$ with $\omega_{\mathrm{eff},\uparrow}=\omega_0+\chi$ and $\omega_{\mathrm{eff},\downarrow}=\omega_0-\chi$. Beyond the vacuum $\ket{0}_{\!\sigma}$, a similar relation also applies to arbitrary Fock states of $H$:
\begin{equation}\label{spin-dep-vaccum}
    \ket*{n}_{\!\sigma}=S(r_{\sigma})\ket{n}.
\end{equation}

We diagonalize the qubit-state-dependent Hamiltonian $H_{\sigma}=\langle \sigma\vert H\vert\sigma\rangle$, giving
\begin{equation}
    H_{\sigma}=(\Delta/2)\langle \sigma\vert \sigma_z\vert\sigma\rangle+\omega_{\sigma}\alpha_{\sigma}^\dagger \alpha_{\sigma}+E_\sigma,
\end{equation}
where $\alpha_{\sigma}=S(r_{\sigma})m S^\dagger(r_{\sigma})$, $\omega_{\sigma}=[\omega_{\mathrm{eff},\sigma}^2-4 \Omega^2]^{1/2}$, and where $E_\sigma=(1/2)(\omega_\sigma-\omega_{\mathrm{eff},\sigma})$ denotes a qubit-state-dependent zero-point energy. The magnetic system is stable when $\omega_{\mathrm{eff},\sigma}^2>4\Omega^2$ for both $\sigma=\uparrow,\downarrow$. In terms of the Fock states $\ket{n}$ of the magnon mode $m$, the ground state $\ket{0}_{\!\sigma}$ is given by $\ket{0}_{\!\sigma}=\sum_n \langle n\vert S(r_\sigma)\vert 0\rangle\ket{n}$. Using Eq.~\eqref{spin-dep-vaccum} to write $\ket{n}$ in the $\alpha_{\bar{\sigma}}$ eigenbasis (i.e.~$\ket{n}=\sum_m \langle m\vert S^\dagger(r_{\bar{\sigma}})\vert n\rangle \ket{m}_{\!\bar{\sigma}}$ with $\bar{\uparrow}=\:\downarrow$), it may then be seen that the ground state of the $\alpha_\sigma$ mode is a squeezed state of the $\alpha_{\bar{\sigma}}$ mode with squeezing parameter $z_\sigma=re^{i\theta_\sigma}$, where $r=r_\downarrow-r_\uparrow>0$ and $\theta_\downarrow=0$ and $\theta_\uparrow=\pi$~\cite{romling2023resolving}:
\begin{equation}\label{squeezing-wrt-up-modes}
\ket{0}_{\!\sigma}=\sum_{n=0}^{\infty} c_{2n}(z_\sigma)\ket{2n}_{\!\bar{\sigma}}.
\end{equation}
Here, $c_{2n}(z_\sigma)=\langle 2n\vert S(z_\sigma)\vert 0\rangle$ are Fock-basis matrix elements of the squeezing operator~\cite{gerry2023introductory}. We stress that the squeezing strength $r$ is tunable via the magnon frequency $\omega_0$, controllable by an external magnetic field. We neglect magnon Kerr-nonlinear effects in Eqs.~\eqref{free-hamiltonian}-\eqref{dispersive-coupling}, which is justified when $\sinh^2 r_\sigma \ll NS$, with $N$ the total number of spins in the magnet. Consequently, the relative squeezing $r$ is constrained by $\sinh^2 r \simeq (2\chi/\Omega)^2 \sinh^2 r_\downarrow \sinh^2 r_\uparrow \ll (NS\chi/\Omega)^2$. Since the spin density is typically on the order of $10^{27}/\mathrm{m}^3$ (as for, e.g., yttrium iron garnet~\cite{sparks1964ferromagnetic}), this condition is not restrictive even for nanoscale magnets. More generally, the requirement can be understood as a condition for truncating the Holstein-Primakoff transformation at leading order in $\langle m^\dagger m\rangle$, which is valid only when the typical number of excitations in the magnet is much smaller than the spin length (in this case, the length ${\sim}NS$ of the macrospin).

\section{Phase-estimation protocol}

The protocol begins with preparation of the qubit and magnon mode in the ground state $\ket{\downarrow}\otimes\ket{0}_\downarrow$, with the qubit and magnon mode coupled by a dispersive coupling yielding a relative squeezing strength of $r\gtrsim 1$. With the qubit initialized in $\ket{\downarrow}$, ground-state initialization can be realized through equilibration by ensuring that $\omega_\downarrow\gg k_B T_E$, where $T_E$ is the temperature of the environment. However, since $\omega_\downarrow\rightarrow 0$ in the limit $r\rightarrow \infty$, the system should first be allowed to equilibrate with $\chi\simeq 0$ ($r\simeq 0$), in which case $\omega_\downarrow\simeq\omega_\uparrow\simeq [\omega_0^2-4\Omega^2]^{1/2}$. When this frequency is on the GHz scale, equilibration to the ground state can then be achieved with temperatures of tens of mK, as may realistically be achieved in a dilution refrigerator. By turning $\chi$ on adiabatically, the strength of the squeezing can subsequently be increased to the desired value while keeping the magnon mode in its ground state $\ket{0}_\downarrow$. The strength of the dispersive coupling could be controlled in-situ by tuning the exchange in the case of spin qubits, or by tuning the qubit frequency in the case of coupling to superconducting qubits~\cite{lachance2017resolving}.

The eigenmode frequency $\omega_\uparrow$ can then be estimated with Heisenberg-limited precision by performing a Ramsey experiment on the qubit. To estimate $\omega_\downarrow$, the qubit could instead be initialized in $\ket{\uparrow}$ so that the magnon mode is prepared in $\ket{0}_{\!\uparrow}$. Concretely, the qubit control needed for the Ramsey sequence consists of an initial Hadamard gate (denoted H in Fig.~\ref{fig:sketch}) at time $t=0$, which prepares the state $\ket{+}=(\ket{\uparrow}+\ket{\downarrow})/\sqrt{2}$. Following a period of free evolution, the qubit is rotated by angle $\varphi$ about the $z$-axis before being measured in the X basis, with the (variable) rotation angle $\varphi$ to be determined adaptively as described below. The idea of using a qubit coupled to a (two-mode) oscillator for quantum metrology was also considered in Ref.~\cite{feng2023robust}; in that work, the protocol required a disentangling operation acting on the joint state of the qubit and oscillator. Here, however, we emphasize that no control over the magnon mode is needed, with all operations performed on the qubit.

As a result of the initial Hadamard, the magnon mode will experience a qubit-conditioned quantum quench: For a qubit in state $\ket{\downarrow}$, evolution is generated by $H_\downarrow$ and the magnon mode remains in its ground state, while for a qubit in state $\ket{\uparrow}$, evolution is instead generated by $H_\uparrow$, for which $\ket{0}_{\!\downarrow}$ is no longer an eigenstate [cf.~Eq.~\eqref{squeezing-wrt-up-modes}]. Since $\ket{0}_{\!\downarrow}$ is a squeezed state of the $\alpha_\uparrow$ mode [cf.~Eq.~\eqref{squeezing-wrt-up-modes}], the squeezed magnons become real (on-shell) conditioned on the qubit being in state $\ket{\uparrow}$, and the different Fock states appearing in the $\alpha_\uparrow$-decomposition of $\ket{0}_{\!\downarrow}$ begin to evolve at multiples of the eigenmode frequency $\omega_\uparrow$. The unitary $U$ describing the evolution of the coupled qubit-magnon system can be written as
\begin{equation}\label{joint-unitary}
    U=\ketbra{\downarrow}\otimes U_\downarrow+\ketbra{\uparrow}\otimes U_\uparrow,
\end{equation}
where $U_\sigma= e^{-i H_\sigma t}$. For a magnon mode in $\ket{0}_{\!\downarrow}$, the $\ket{\downarrow}$-conditioned evolution generated by $H_\downarrow$ acts trivially up to a phase: $U_\downarrow\ket{0}_{\!\downarrow}\propto\ket{0}_{\!\downarrow}$. The nontrivial evolution generated by $U_\uparrow$ can then be probed through a measurement of the qubit coherence.

Following the free-evolution window, the qubit is rotated according to $Z_\varphi=e^{-i\varphi(t)\sigma_z/2}$, with $\varphi(t)$ to be determined in the manner described below. The qubit is then measured in the X basis, yielding the expectation value $\langle \sigma_x\rangle=\mathrm{Tr}\{Z_\varphi U\rho_0 U^\dagger Z_\varphi^\dagger \sigma_x\}$, where $\rho_0=\ketbra{\psi_0}$ for $\ket{\psi_0}=\ket{+}\otimes \ket{0}_\downarrow$. For a total free-evolution time $t$, this expectation value can be written as 
\begin{equation}\label{X-average}
    \langle \sigma_x\rangle_t=\Re\left\{e^{-i\theta(t)}{}_{\downarrow\!}\langle 0 \vert e^{-i\omega_\uparrow\alpha_\uparrow^\dagger\alpha_\uparrow t} \vert 0\rangle_{\!\downarrow}\right\},
\end{equation}
where $_{\downarrow\!}\langle 0 \vert e^{-i\omega_\uparrow\alpha_\uparrow^\dagger\alpha_\uparrow t} \vert 0\rangle_{\!\downarrow}$ is the dynamical overlap of the initial and time-evolved states of the magnon mode, and where $\theta(t)=\varphi(t)-(\Delta+\delta_{\mathrm{zp}})t$ with $\delta_{\mathrm{zp}}=E_\uparrow-E_\downarrow$ giving the difference in zero-point energy of the magnon mode depending on the qubit state.  

Since $\ket{0}_{\!\downarrow}$ contains only even Fock states of the $\alpha_\uparrow$ mode [cf.~Eq.~\eqref{squeezing-wrt-up-modes}], the dynamical overlap is periodic, returning to its initial value of 1 at multiples of $\pi/\omega_\uparrow$. For $\theta(t)=0$, $\langle \sigma_x\rangle_t$ will therefore exhibit periodic collapses and recurrences occurring at multiples of $\pi/\omega_\uparrow$, which become increasingly localized as the relative squeezing strength $r=r_\downarrow-r_\uparrow$ is increased (Fig.~\ref{fig:revivals-information}). The increase in steepness with $r$ can be understood through the lens of the orthogonality catastrophe as being a consequence of the magnon mode having a quantum speed limit that scales extensively with the average mode occupation~\cite{fogarty2020orthogonality}: Since the Bures angle $\Theta(t)=\arccos{\lvert_{\downarrow}\!\langle 0 \vert U_\uparrow(t) \vert 0\rangle_{\!\downarrow}\rvert}$ between the initial and time-evolved states is bounded by the quantum speed limit~\cite{wootters1981statistical,taddei2013quantum}, any quantities depending on $\Theta(t)$, including, for instance, the qubit coherence $\lvert \langle\sigma_+\rangle_t\rvert=(1/2)\cos{\Theta(t)}$, will consequently evolve faster with increasing $r$ (or equivalently, faster with increasing $\bar{n}=\sinh^2{r}$, where $\bar{n}$ is the average number of magnons produced by the qubit-conditioned quantum quench). 

For $\theta(t)\neq 0$, the recurrences in $\langle \sigma_x\rangle_t$ may be modulated by a beating pattern. Such beating would ultimately reduce the Fisher information calculated below (as may be verified numerically). The beating can be removed by adaptively tuning $\varphi(t)$ until a pattern of periodic recurrences with height 1 is recovered \footnote{For $\varphi(t)=\lambda T_g(t)$, where $\lambda$ has dimensions of frequency and $T_g(t)=c't$, $c'\in\mathbbm{R}$, is a gate time that can be varied, the beating is removed when $c'=c $ with $c=(\Delta+\delta_{\mathrm{zp}})/\lambda$. For some initial setting $T_g(t)=c't$, the constant $c'$ can then be updated adaptively based on measured experimental data until this condition is satisfied. (In practice, the $Z$ rotation could be implemented virtually.) In the event that such control is unavailable, full-amplitude recurrences at multiples of $\pi/\omega_\uparrow$ could alternatively be obtained from $\langle\sigma_x\rangle_t^2+\langle\sigma_y\rangle_t^2$. Maximum-likelihood estimation performed in the vicinity of such a recurrence could be used to obtain an estimate of $\phi$ with a standard deviation scaling inversely in the number of magnons. The scaling provided by this approach is therefore Heisenberg-limited as well, though with a smaller overall prefactor.}. We therefore set $\theta(t)=0$ from hereon out under the assumption that $\Delta+\delta_{\mathrm{zp}}$ has been learned via this procedure, allowing it to be deterministically compensated.

\section{Heisenberg-limited scaling near the post-quench recurrences}
With $\varphi(t)=(\Delta+\delta_{\mathrm{zp}})t$ (mod $2\pi$), the probability $p(+\vert\phi)$ of measuring the qubit in state $\ket{+}$ given some value of $\phi=\omega_\uparrow t$ follows from Eq.~\eqref{X-average} and is given by
\begin{equation}\label{probability-qubit-measurement}
    p(+\vert\phi)=\frac{1}{2}+\frac{1}{2}\sum_n \lvert c_{2n}(r)\rvert^2 \cos{(2n\phi)}.
\end{equation}
The Heisenberg-limited scaling enabled by this protocol in the vicinity of a qubit-coherence recurrence (occurring at times $t$ such that $\phi =\ell\pi$ with $\ell\in\mathbb{Z}$) ultimately arises from super-resolution $\phi\mapsto 2n\phi$~\cite{resch2007time} of the parameter $\phi$ in Eq.~\eqref{probability-qubit-measurement}, as we now show by calculating the Fisher information associated with the qubit measurement.

\begin{figure}
    \centering
    \includegraphics[width=\linewidth]{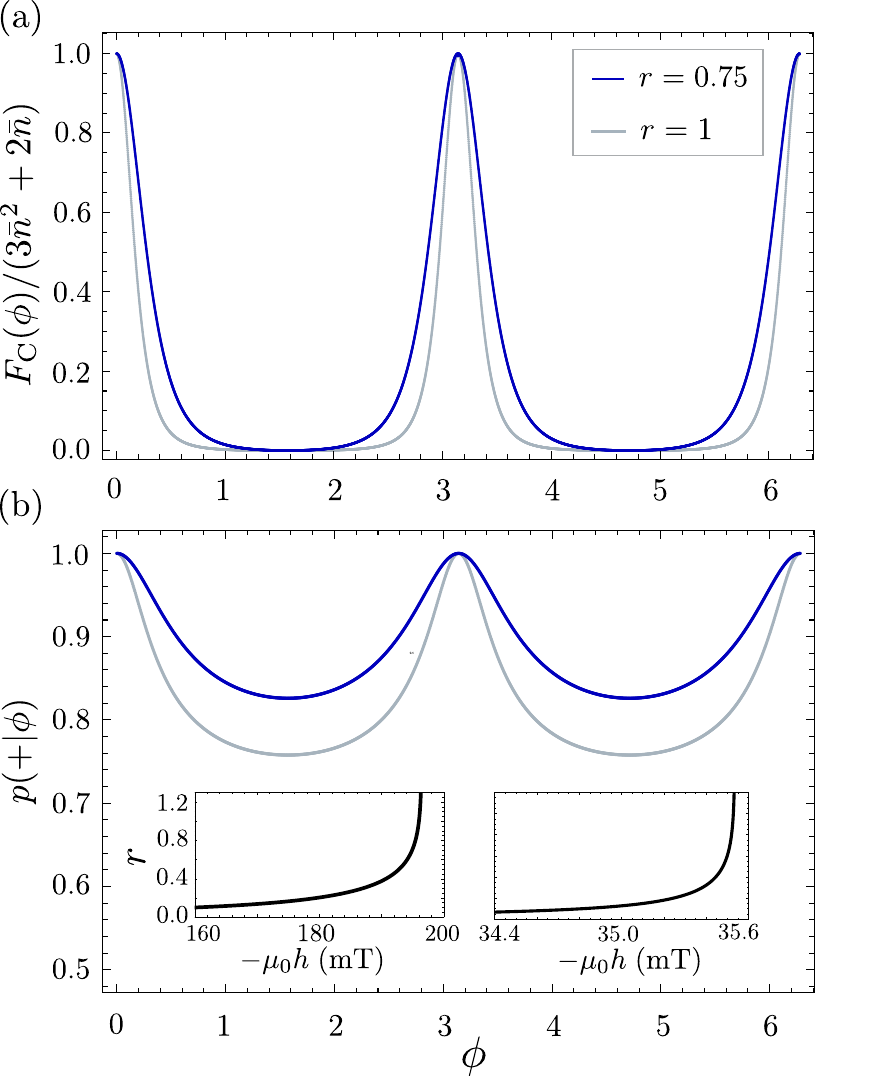}
    \caption{(a) Fisher information $F_{\mathrm{C}}(\phi)$ evaluated using Eqs.~\eqref{probability-qubit-measurement} and \eqref{fisher-information-definition} for $r=0.75$ (dark blue) and $r=1$ (light blue). As described in the text, the duration $T$ of the free-evolution window should be chosen so that $\phi=\omega_\uparrow T$ is approximately equal to $\ell\pi$. The free-evolution time $T$ could be tweaked adaptively using the tunability of the squeezing parameter as a function of the applied magnetic field. (b) Probability $p(+\vert\phi=\omega_\uparrow t)=(1/2)(1+\langle \sigma_x\rangle_t)$ of measuring the qubit in the state $\ket{+}$ given a value of $\phi$. In the absence of ground-state squeezing ($\Omega=0$, corresponding to $r_\uparrow=r_\downarrow=r=0$), the probability is $p(+\vert \phi)=1$ independent of $\phi$, and the Fisher information associated with the measurement is zero. Left inset: Squeezing parameter $r$ as a function of the applied magnetic field $h$ for $(2SK_z-SK_y)/2\pi=7$ GHz and $\Omega/2\pi=\chi/2\pi=0.5$ GHz. A field of 180 mT corresponds to a Zeeman splitting of 5.04 GHz (assuming a $g$-factor of 2). Right inset: The same range of squeezing parameters could be achieved with the same value of $\Omega/2\pi = 0.5$ GHz, $(2SK_z-SK_y)/2\pi=2$ GHz, and a considerably smaller dispersive coupling of $\chi/2\pi = 5$ MHz~\cite{weng2026magnon}.}
    \label{fig:revivals-information}
\end{figure}

The best precision that can be achieved with $M$ independent measurements and an unbiased estimator of $\phi$ is given by the Cram\'er-Rao bound~\cite{cramer1999mathematical},
$\delta\phi^2\geq [MF_{\mathrm{C}}(\phi)]^{-1},$
where here, $F_{\mathrm{C}}(\phi)$ is the Fisher information quantifying the amount of information about $\phi$ that can be extracted from the probability distribution describing the measurement outcomes. For a binary measurement like the X-basis qubit measurement considered here,
\begin{equation}\label{fisher-information-definition}
F_{\mathrm{C}}(\phi)=\sum_{x=\pm}p(x\vert\phi)\left[\partial_\phi\mathrm{ln}\:p(x\vert\phi)\right]^2.
\end{equation}
This quantity can be evaluated using the expression for $p(+\vert\phi)$ given in Eq.~\eqref{probability-qubit-measurement}. For $\vert \phi-\ell\pi\rvert\ll \bar{n}^{-1}$ with $\bar{n}\gtrsim 1$, we can expand the trigonometric functions to leading nontrivial order in $\phi$, 
giving $F_{\mathrm{C}}(\phi)\simeq \sum_n\lvert c_{2n}\rvert^2(2n)^2$. This sum is equal to the average of $(\alpha_\uparrow^\dagger\alpha_\uparrow)^2$ with respect to $\ket{0}_{\!\downarrow}$ [cf.~Eq.~\eqref{squeezing-wrt-up-modes}]. Using Wick's theorem, this average can then be rewritten entirely in terms of the expectation values ${}_{\!\downarrow}\!\langle 0\vert \alpha_\uparrow^2\vert 0\rangle_{\!\downarrow}=-(1/2)\sinh{(2r)}$ and $_{\downarrow}\!\langle 0\vert \alpha_\uparrow^\dagger\alpha_\uparrow\vert 0\rangle_\downarrow=\sinh^2{r}$, giving 
\begin{equation}\label{local-max-fisher}
    F_{\mathrm{C}}(\phi)\simeq 3\bar{n}^2+2\bar{n}, \quad \vert\phi-\ell\pi\rvert\ll \bar{n}^{-1},\quad\bar{n}\gtrsim 1,
\end{equation}
where $\bar{n}=\sinh^2{r}$ is the average number of real magnons produced by the quantum quench. The Heisenberg scaling $\bar{n}^2$ in Eq.~\eqref{local-max-fisher} is the main result of this work. Since $\bar{n}=0$ in the absence of squeezing terms $\propto\Omega$ in $H_0$ [Eq.~\eqref{free-hamiltonian}], the parameter estimation enabled by this protocol is due entirely to the quantum features of the magnon state. Although Eq.~\eqref{local-max-fisher} gives the scaling of $F_{\mathrm{C}}(\phi)$ with $\bar{n}$ for $\bar{n}\gtrsim 1$, the fact that $F_{\mathrm{C}}(\phi)=0$ for $\bar{n}=\sinh^2{r}=0$ follows directly from the fact that for $r=0$, we have $p(+\vert\phi)=1$ independent of $\phi$. More concretely, this is because for $\Omega=0$ (leading to $r=0$), the magnon eigenbasis is independent of the state of the qubit, with $\ket{0}$ giving the ground state of the magnon mode independent of the qubit. The state $\ket{+}\ket{0}$ is then an eigenstate of the (rotating-frame) Hamiltonian $H$ given in Eqs.~\eqref{free-hamiltonian}-\eqref{dispersive-coupling}, and as a result, no quantum quench occurs when the qubit is rotated from $\ket{\downarrow}$ to $\ket{+}$ via the Hadamard gate. 

Away from the regime $\vert\phi-\ell\pi\rvert\ll \bar{n}^{-1}$, $F_{\mathrm{C}}(\phi)$ can be evaluated numerically [Fig.~\ref{fig:revivals-information}(a)]. Since maximum-likelihood estimation saturates the Cram\'er-Rao bound in the limit of many measurements~\cite{cramer1999mathematical}, Heisenberg-limited estimation of the magnon-mode frequency $\omega_\uparrow$ can be achieved by performing maximum likelihood estimation of $\phi$ based on the probability distribution $p(+\vert \phi)$ in the vicinity of any post-quench recurrence of the qubit coherence [Fig.~\ref{fig:revivals-information}(b)]. Concretely, the protocol would require timing the total duration $T$ of the free-evolution window so that $\phi$ is close to $\ell\pi$, where the dependence of the qubit-measurement outcomes is strongly dependent on $\phi$. This tuning of $T$ could be performed adaptively by controlling the squeezing parameter $r$ via the external magnetic field $h$ [Fig.~\ref{fig:revivals-information}(b,inset)], beginning with a low value of $r$ and increasing it as the location of the periodic recurrences is revealed. 

In the left inset of Fig.~\ref{fig:revivals-information}(b), we plot the squeezing parameter $r=r_\downarrow-r_\uparrow$ as a function of the applied magnetic field $\mu_0 h$ in mT, with $\omega_0/2\pi=(7+\vert\gamma\vert\mu_0 h)$ GHz and $\chi/2\pi=\Omega/2\pi=0.5$ GHz chosen so that the range of applied magnetic fields over which $r\approx 1$ gives Zeeman splittings of approximately 5 GHz, assuming a $g$-factor of 2. In the right inset, we assume a much smaller dispersive coupling of $\chi/2\pi=5$ MHz, as has recently been realized for cavity-mediated coupling between a superconducting transmon qubit and yttrium iron garnet (YIG) sphere~\cite{weng2026magnon}. In this case, the same range of squeezing strengths could be realized with $\Omega/2\pi=0.5$ GHz and $\omega_0/2\pi=(2+\vert \gamma\vert \mu_0 h)$ GHz. More generally, for values of $\chi\ll \omega_0$, the squeezing parameter $r$ is given by $r=(2\Omega/\omega_0)/[(2\Omega/\omega_0)^2-1](\chi/\omega_0)+O((\chi/\omega_0)^2)$. Hence, a value $r\gtrsim 1$ can be generated even for $\chi/\omega_0\ll 1$ provided $2\Omega$ is itself comparable to $\omega_0$, as could be realized by taking advantage of the tunability of $\omega_0$ via the applied magnetic field. For values of $r$ close to one, the advantage coming from the Heisenberg scaling is modest but nevertheless exceeds the precision associated with the SQL; for $r=1$, for instance, the Fisher information per magnon is $F_\mathrm{C}(\phi)/\bar{n}\simeq 6$ times larger than would be achieved with SQL scaling.

\begin{figure}
    \centering
    \includegraphics[width=\linewidth]{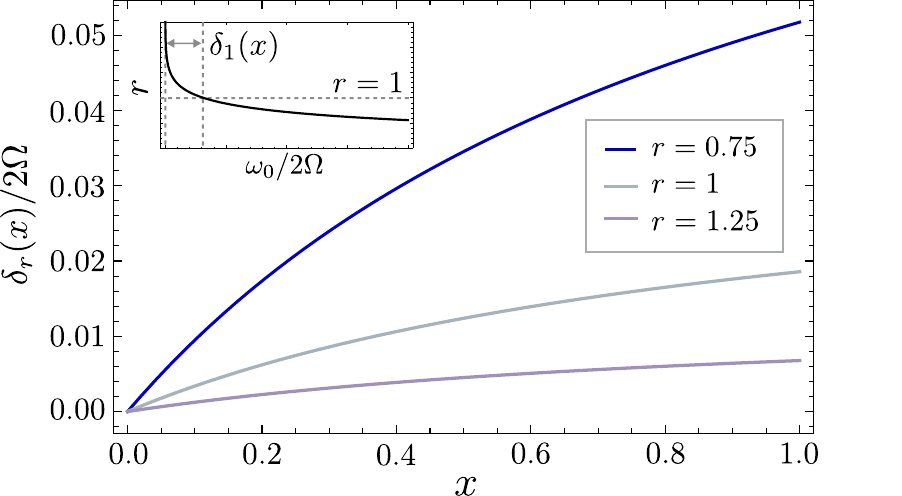}
    \caption{Difference $\delta_r(x)$ [Eq.~\eqref{stability-req}] between the magnon-mode frequency $\omega_0$ required to achieve a given value of $r$ at fixed $x=\chi/2\Omega$ and the frequency $2\Omega+\chi$ at which the system becomes unstable. Inset: Schematic illustrating the definition of $\delta_r(x)$ for $r=1$.}
    \label{fig:field-stability}
\end{figure}
It is worth remarking that the requirements on the stability of the applied field $h$ become more stringent as $r$ is increased. In particular, for values of $\omega_0(h)$ below $2\Omega+\chi$, the Hamiltonian description given in Eqs.~\eqref{free-hamiltonian}-\eqref{dispersive-coupling} breaks down as the system becomes unstable; this is indicated by a divergence of $r$. In Fig.~\ref{fig:field-stability}, we plot the frequency difference $\delta_r(x=\chi/2\Omega)$ between the critical value of $\omega_0$ at which the system becomes unstable, and the value of $\omega_0$ required to achieve a particular value of $r$ at fixed $x=\chi/2\Omega$:
\begin{equation}\label{stability-req}
    \frac{\delta_r(x)}{2\Omega}=\sqrt{x^2+1+2x\coth{2r}}-(1+x).
\end{equation}
In the presence of a variable field $h+\delta h$, where $\delta h$ denotes a small variation about the average field $h$, the standard deviation $\sigma_h$ of $\delta h$ should consequently satisfy $\vert \gamma\vert \mu_0\sigma_h\ll 2\Omega \delta_r(\chi/2\Omega)$ to ensure stability of the magnetic system. For the scenario illustrated in the left inset of Fig.~\ref{fig:revivals-information}(b), for instance, we have $2\Omega \delta_r(\chi/2\Omega)\approx 12$ MHz for $r=1$. For  $\vert \gamma\vert \mu_0\sigma_h=1$ MHz, a plot of the field-averaged probability $p(+\vert \phi)=\int d(\delta h)\:p(\delta h)p(+\vert\phi(\delta h))$ does not visibly differ from that shown in Fig.~\ref{fig:revivals-information}(b) when $p(\delta h)$ is taken to be normally distributed with mean zero and standard deviation $\sigma_h$. This indicates that the Fisher information is quite robust to  variations in the applied field provided $\vert \gamma\vert \mu_0\sigma_h\ll 2\Omega \delta_r(\chi/2\Omega)$. 

\section{Effects of decoherence}

To calculate the effects of qubit dephasing on the Fisher information $F_{\mathrm{C}}$, we now include in $H_0$ an additional term $\eta\sigma_z/2$, where $\eta$ is a zero-mean, Gaussian random variable with $\llangle \eta\rrangle=0$ and $\llangle \eta^2\rrangle=2/T_2^*$. Here, $T_2^*$ is the qubit dephasing time and $\llangle \cdot \rrangle$ denotes an average over the probability distribution $p(\eta)$ governing the distribution of $\eta$-values from one shot to the next: $\llangle \cdot\rrangle=\int d\eta \:p(\eta)(\cdot)$. For a fixed value of $\eta$, $\langle \sigma_x\rangle_t=(1/2)e^{i\eta t}{}_{\downarrow}\!\langle 0\vert U_\uparrow\vert 0\rangle_\downarrow+\mathrm{c.c.}$, giving
\begin{equation}
    \llangle \langle \sigma_x\rangle_t\rrangle=e^{-(t/T_2^*)^2}\Re_{\downarrow}\!\langle 0\vert U_\uparrow(t)\vert 0\rangle_\downarrow.
\end{equation}
Expanding as before about $\phi=\ell\pi$, we find that in the presence of qubit dephasing, the local maxima in the Fisher information will be modified by a decaying Gaussian envelope according to 
\begin{equation}
    F_{\mathrm{C}}(\phi)\simeq e^{-\left(K \ell\right)^2}\left(3\bar{n}^2+2\bar{n}\right),
\end{equation}
valid for $\lvert \phi-\ell\pi\vert\ll \bar{n}^{-1}$. Here, we have introduced a dimensionless parameter $K=\pi/(\omega_\uparrow T_2^*)$ controlling the decay of the $m^{\mathrm{th}}$ recurrence relative to the ideal case where $T_2^*\rightarrow\infty$.

Hence, provided $\omega_\uparrow T_2^*\gg 1$, qubit dephasing should not present a limiting error source provided maximum-likelihood estimation is carried out in the vicinity of an early revival where $e^{-(K\ell)^2}\approx 1$. Magnon decoherence is likewise negligible when the recurrence time $\ell\pi/\omega_\uparrow$ is short compared to the magnon coherence time $T_2^{\rm mag}$. For GHz-scale magnon frequencies, $\pi/\omega_\uparrow$ is on the order of a few nanoseconds, while experiments on YIG report $T_2^{\rm mag}\sim0.1$--$1\,\mu{\rm s}$~\cite{tabuchi2015coherent,lachance2017resolving,lachance2020entanglement,xu2023quantum,weng2026magnon}.

\section{Conclusion} In this work, we have presented a measurement protocol enabling Heisenberg-limited metrology with equilibrium ground-state squeezing in magnetic systems. This strategy can be used to estimate an eigenmode frequency for the purpose of system characterization.
In the proposed protocol, dispersive coupling to a qubit can be used to subject the magnon mode to a quantum quench conditioned on the state of the qubit. This quench produces real squeezed magnons in a process that may be viewed as a magnon analog of the dynamical Casimir effect. The interaction of the qubit with the magnon system then manifests as super-resolution of an unknown parameter in the qubit coherence, enabling precision beyond the SQL. 
In the future, the protocol presented here could inspire novel ways of exploiting the quantum properties of magnetic systems for tasks beyond the reach of classical states.

\begin{acknowledgments}
ZMM thanks W.~A.~Coish for useful discussions. This work was supported by the Georg H. Endress Foundation, by the Swiss National Science Foundation, NCCR SPIN (grant no. 51NF40-180604), and by the Ibn Sina Global Scholarship. DL acknowledges the Deanship of Research at KFUPM and the Quantum Center for the support received under Grant no. CUP25102 and no. INQC2500.
\end{acknowledgments}

\appendix

\section{Derivation of the NV-center--magnon Hamiltonian}\label{app:NV}

In this appendix, we derive the strength of the NV-center--magnon coupling obtained by quantizing the magnet's stray field. An NV-center spin (denoted $\bm{S}_{\mathrm{NV}}$) will interact with the stray field of a magnet via a Zeeman coupling,
\begin{equation}
    H_{\mathrm{Z}}=\gamma\bm{S}_{\mathrm{NV}}\cdot\bm{B}_{\mathrm{tot}}(\bm{r}),
\end{equation}
where $\gamma=28$ GHz/T is the electron gyromagnetic ratio and $\bm{B}_{\mathrm{tot}}(\bm{r})$ is the total magnetic field at the location $\bm{r}$ of the NV center. This magnetic field is the sum of the external (applied) field $\bm{B}_{\mathrm{ext}}=B_{\mathrm{ext}}\hat{z}$ and the stray field $\bm{B}_{\mathrm{stray}}(\bm{r})$ due to the magnet itself.

To calculate $\bm{B}_{\mathrm{stray}}(\bm{r})$, we start from the vector potential $\bm{A}_{\mathrm{stray}}(\bm{r})$, in terms of which $\bm{B}_{\mathrm{stray}}=\nabla\times\bm{A}_{\mathrm{stray}}$. This vector potential can be calculated as an integral over the magnet given by~\cite{griffiths2023introduction}
\begin{equation}\label{vector-potential}
    \bm{A}_{\mathrm{stray}}(\bm{r})=\frac{\mu_0}{4\pi}\int_{V}dV'\frac{\bm{J}(\bm{r')}}{\vert\bm{r}-\bm{r'}\vert}+\frac{\mu_0}{4\pi}\oint_S dS'\frac{\bm{K}(\bm{r'})}{\vert \bm{r}-\bm{r}'\rvert},
\end{equation}
where $\bm{J}=\nabla\times\bm{M}$ and $\bm{K}=\bm{M}\times \hat{\bm{n}}$ are the bound and surface charges due to the magnetization $\bm{M}$ of the magnet. Here, $\int_V$ and $\oint_S$ are integrals over the bulk and surface of the magnet, respectively. 

We assume the magnet is sufficiently small for only the Kittel mode to be relevant. To relate the magnetization $\bm{M}$ to Kittel-mode operators, we begin by writing $\bm{M}=\gamma V^{-1}\bm{S}_{\mathrm{tot}}$, where $V$ is the volume of the magnet and $\bm{S}_{\mathrm{tot}}$ is the collective spin. Due to the spatial uniformity of the Kittel mode, we retain only the contribution coming from the surface term in Eq.~\eqref{vector-potential}, giving
\begin{equation}\label{stray-field}
    \bm{B}_{\mathrm{stray}}(\bm{r})=\frac{\mu_0}{4\pi}\nabla\times\oint_S dS'\frac{\bm{M}\times \hat{\bm{n}}'}{\vert \bm{r}-\bm{r'}\vert}.
\end{equation}

Since the curl is with respect to the observation coordinate $\bm{r}$, it can be moved inside the surface integral. To simplify, we can then use the vector identity $\nabla\times (\bm{A}\times \bm{B})=(\bm{B}\cdot \nabla)\bm{A}-\bm{B}(\nabla\cdot \bm{A})-(\bm{A}\cdot \nabla)\bm{B}+\bm{A}(\nabla\cdot \bm{B})$ with $\bm{A}=\bm{M}$ and $\bm{B}=\hat{\bm{n}}'$. Since $\bm{M}$ is spatially uniform, we have $\nabla \bm{M}=0$ and $\nabla\cdot \bm{M}=0$, giving 
\begin{align}
\begin{aligned}\label{vector-identity}
    \nabla\times(\bm{M}\times \hat{\bm{n}}')&=-(\bm{M}\cdot \nabla)\hat{\bm{n}}'+\bm{M}(\nabla\cdot \hat{\bm{n}}')\\
    &=-\nabla \hat{\bm{n}}'\cdot\bm{M}+\bm{M}(\nabla\cdot\hat{\bm{n}}').
\end{aligned}
\end{align}
This expression makes it clear that there are two qualitatively different terms in the stray field. The first points along a direction set by $\bm{M}$ and originates from the term $(\nabla\cdot\hat{\bm{n}}')\bm{M}$, while the second ultimately couples different components of $\bm{S}_{\mathrm{NV}}$ and $\bm{S}_{\mathrm{tot}}$ in a manner controlled by $\nabla\hat{\bm{n}}'$. Substituting Eq.~\eqref{vector-identity} back into Eq.~\eqref{stray-field}, we can then write the Zeeman interaction as
\begin{equation}
    H_{\mathrm{Z}}=\gamma B_{\mathrm{ext}}S_{\mathrm{NV}}^z+\gamma B(\bm{r})\bm{S}_{\mathrm{NV}}\cdot\bm{S}_{\mathrm{tot}}-\gamma \bm{S}_{\mathrm{NV}}\cdot\bm{H}(\bm{r})\cdot \bm{S}_{\mathrm{tot}},
\end{equation}
where
\begin{align}
    &B(\bm{r})=\frac{\mu_0\gamma}{4\pi V}\oint_{S}dS' \frac{\nabla\cdot\hat{\bm{n}}'}{\vert \bm{r}-\bm{r}'\vert},\label{scalar}\\
    &\bm{H}(\bm{r})=\frac{\mu_0\gamma }{4\pi V}\oint_{S}dS'\frac{\nabla\hat{\bm{n}}'}{\vert\bm{r}-\bm{r}'\vert}.\label{tensor}
\end{align}
If we assume that the equilibrium direction of $\bm{S}_{\mathrm{tot}}$ is predominantly along the external field $\bm{B}_{\mathrm{ext}}$, then we can use a Holstein-Primakoff transformation to write $\bm{S}_{\mathrm{tot}}$ in terms of magnon operators with
\begin{align}
    &S_{\mathrm{tot}}^z= S_{\mathrm{tot}}-m^\dagger m,\label{sx}\\
    &S_{\mathrm{tot}}^y\simeq i\sqrt{\frac{S_{\mathrm{tot}}}{2}}(m^\dagger-m),\label{sy}\\
    &S_{\mathrm{tot}}^x\simeq \sqrt{\frac{S_{\mathrm{tot}}}{2}}(m^\dagger+m),\label{sz}
\end{align}
valid up to corrections $O(1/S_{\mathrm{tot}})=O(V^{-1})$.

We next write $H_{\mathrm{Z}}$ as the sum of a classical-field contribution $H_{\mathrm{Q}}$ and an interaction Hamiltonian $H_{\mathrm{int}}$ that couples spin and magnon operators:
\begin{equation}\label{zeeman-2}
    H_{\mathrm{Z}}= \underbrace{\gamma  \bm{B}_{\mathrm{cl}}(\bm{r})\cdot \bm{S}_{\mathrm{NV}}}_{H_{\mathrm{Q}}}+H_{\mathrm{int}}+O(1/S_{\mathrm{tot}}),
\end{equation}
where
\begin{equation}\label{classical-field}
    \bm{B}_{\mathrm{cl}}(\bm{r})= \left(B_{\mathrm{ext}} +S_{\mathrm{tot}}B(\bm{r})\right)\hat{z}-\sum_{j=x,y,z}S_{\mathrm{tot}}H_{j z}(\bm{r})\hat{j}.
\end{equation}
The NV-magnon coupling $H_{\mathrm{int}}$ can be written as
\begin{align}\label{interaction-qubit-magnon}
H_{\mathrm{int}}=g_y\hat{\Sigma}_y(\bm{r})\hat{P}+g_x\hat{\Sigma}_x(\bm{r})\hat{X},
\end{align}
where $\hat{X}=(m+m^\dagger)/\sqrt{2}$ and  $\hat{P}=i(m-m^\dagger)/\sqrt{2}$ are conjugate position and momentum operators for the Kittel mode, $\hat{\Sigma}_\nu$ ($\nu=x,y)$ are NV-center spin operators, and where $g_\nu$ are coupling constants:
\begin{align}
    &\hat{\Sigma}_\nu(\bm{r})=\sum_{\mu=x,y,z}\sigma_\mu \frac{g_{\mu \nu}(\bm{r})}{g_\nu(\bm{r})},\label{linear-combination}\\
    &g_\nu(\bm{r})=\left[\sum_{\mu=x,y,z}g_{\mu \nu}^2(\bm{r})\right]^{1/2},\\
    &g_{\mu \nu}(\bm{r})=\gamma\sqrt{S_{\mathrm{tot}}}\left[H_{\mu \nu}(\bm{r})-\delta_{\mu,\nu}B(\bm{r})\right].\label{coupling-coefficients}
\end{align}
In Eq.~\eqref{interaction-qubit-magnon}, we have neglected a coupling $\propto m^\dagger m$ since it is $S_{\mathrm{tot}}$-independent and consequently much weaker than terms $\propto \sqrt{S_{\mathrm{tot}}}$ for $S_{\mathrm{tot}}\gg 1$~\cite{dols2024magnon}. 

For a point $\bm{r}$ well outside the magnet,  $B(\bm{r})$ and $\bm{H}(\bm{r})$ [Eqs.~\eqref{scalar} and \eqref{tensor}] are independent of the volume of the magnet. This can be seen from a simple argument based on the multipole expansion
\begin{equation}
    \frac{1}{\vert\bm{r}-\bm{r}'\vert}=\frac{1}{r}\sum_{n=0}^{\infty} \left(\frac{r'}{r}\right)^nP_n(\cos{\alpha}),
\end{equation}
where $r$ and $r'$ denote the distances to the observation and integration points, respectively, $P_n$ is the $n^{\mathrm{th}}$ Legendre polynomial, and where $\alpha$ is the angle subtending $\bm{r}$ and $\bm{r}'$. In a multipole expansion of a magnetic field, the monopole ($n=0$) term vanishes, leaving the dipole ($n=1$) term as the first nontrivial contribution. Writing the surface element $dS'=r' dr'\: d\theta$ in polar coordinates, it then follows that the dipole term scales like the volume $V$ of the integration area. The factor of $V$ in the denominator of Eqs.~\eqref{scalar}-\eqref{tensor} cancels out this dependence. We therefore keep track of factors of $V$ by keeping track of factors of $S_{\mathrm{tot}}$, as we have been doing so far.

\subsubsection{Stray field in the dipole approximation}

For simplicity, we treat the magnet as a small particle with magnetic moment $\bm{m}=\bm{M}V=\gamma \hbar \bm{S}_{\mathrm{tot}}$, where we have reintroduced an explicit dependence on $\hbar$ for the purpose of the numerical estimate below. In this case, the stray field at a position $\bm{r}=r\hat{\bm{r}}$ is given by
\begin{equation}\label{point-particle}
    \bm{B}_{\mathrm{stray}}(\bm{r})=\frac{\mu_0 \gamma\hbar}{4\pi r^3}\left[3(\bm{S}_{\mathrm{tot}}\cdot\hat
    {\bm{r}})\hat{\bm{r}}-\bm{S}_{\mathrm{tot}}\right].
\end{equation}
We write the unit vector as $\hat{\bm{r}}=n_x \hat{\bm{x}}+n_y\hat{\bm{y}}+n_z\hat{\bm{z}}$. Substituting this expression into Eq.~\eqref{point-particle} with $\bm{S}_{\mathrm{tot}}=(S_x,S_y,S_z)$ gives
\begin{equation}\label{stray-dipole}
    \bm{B}_{\mathrm{stray}}(\bm{r})=\frac{\mu_0\gamma\hbar}{4\pi r^3}\left(\begin{matrix}
        (3n_x^2-1) S_x+3n_x n_y S_y+3n_xn_z S_z\\
        3n_y n_x S_x+(3n_y^2-1)S_y+3n_yn_zS_z\\
        3n_zn_x S_x+3n_zn_y S_y+(3n_z^2-1)S_z
    \end{matrix}\right).
\end{equation}
As before, we can write the stray field in terms of Kittel-mode magnon operators through a Holstein-Primakoff transformation [Eqs.~\eqref{sx}-\eqref{sz}] to obtain the classical field $\bm{B}_{\mathrm{cl}}(\bm{r})$ [Eq.~\eqref{classical-field}] and NV-magnon coupling $H_{\mathrm{int}}$ [Eq.~\eqref{interaction-qubit-magnon}]. For the classical field, we obtain
\begin{align}\label{classical-field-example}
\begin{aligned}
    \bm{B}_{\mathrm{cl}}(\bm{r})&=B_{\mathrm{ext}}\hat{z}+\frac{\mu_0\gamma\hbar}{4\pi r^3}(3n_z^2-1)S_{\mathrm{tot}}\hat{z}\\&+\frac{\mu_0\gamma\hbar}{4\pi r^3}3n_zn_y S_{\mathrm{tot}}\hat{y}+\frac{\mu_0\gamma\hbar}{4\pi r^3}3n_zn_x S_{\mathrm{tot}}\hat{x},
\end{aligned}
\end{align}
from which we can directly identify the quantities appearing in the general formula given in Eq.~\eqref{classical-field}:
\begin{align}
    B(\bm{r})&=-\frac{\mu_0\gamma\hbar}{4\pi r^3},\\
    H_{\alpha \beta}(\bm{r})&=-\frac{\mu_0\gamma\hbar}{4\pi r^3}3n_\alpha n_\beta.
\end{align}
These quantities also appear as expected in the spin-magnon interaction $H_{\mathrm{int}}=g_y\hat{\Sigma}_y\hat{P}+g_x\hat{\Sigma}_x\hat{X}$ obtained directly from Eq.~\eqref{stray-dipole} [cf.~Eq.~\eqref{interaction-qubit-magnon}], where explicitly, we have
\begin{align}
    g_\nu\hat{\Sigma}_\nu=\frac{\gamma}{\hbar}\sqrt{S_{\mathrm{tot}}}\frac{\mu_0\gamma\hbar}{4\pi r^3}\left[S_{\mathrm{NV}}^\nu-\sum_{\alpha=x,y,z}3n_\alpha n_\nu S_{\mathrm{NV}}^\alpha\right].
\end{align}

We assume for concreteness that the NV center is located along the $\hat{z}$ axis, so that $n_z=1$ and $n_{y}=n_z=0$.  The NV-magnon interaction Hamiltonian is then given by
\begin{align}
    H_{\mathrm{int}}&=\sqrt{2}ig S_{\mathrm{NV}}^y(m^\dagger -m)-\sqrt{2}gS_{\mathrm{NV}}^x(m+m^\dagger)\\
    &=-\sqrt{2}g(S_{\mathrm{NV}}^+ m+S_{\mathrm{NV}}^- m^\dagger),
\end{align}
where
\begin{equation}
    g(r)\equiv \frac{\gamma}{\hbar}\sqrt{S_{\mathrm{tot}}}\frac{\mu_0\gamma\hbar}{4\pi r^3}.
\end{equation}
In the dispersive regime (where the NV-magnon detuning exceeds $g$), an interaction of the form $S_{\mathrm{NV}}^+m+\mathrm{h.c.}$ will produce the required dispersive coupling assumed in the paper.

We now estimate the coupling strength $g(r)$. Grouping parameters, we have $\mu_0\gamma\hbar/4\pi\approx 1.9 \times 10^{-30}\: \mathrm{T}\cdot m^3$ and $\gamma/\hbar\approx 1.8\times 10^{11}\:\frac{\mathrm{Hz}}{\mathrm{T}}$. The total macrospin $S_{\mathrm{tot}}= \rho_s S V$ depends on both the spin density $\rho_s$, on-site spin $S$, and volume $V$ of the magnet. For YIG, we have $\rho_s\approx 10^{27}\:/m^3$~\cite{sparks1964ferromagnetic}, $S=5/2$, and we take $V\approx a^3\: m^3$ to scale like the cube of the typical size $a$ of the magnet in meters so that $\sqrt{S_{\mathrm{tot}}}\approx 10^{27/2}\sqrt{ a^3 }$.  Since $r$ must necessarily correspond to a point outside the magnet, we write $r=a(1+x)\:m$ where $x>0$ is the distance from the surface of the magnet to the location of the spin, expressed in units of $a$. The combination of length scales controlling $g(r)$ is therefore
\begin{equation}
    g(r)\approx \frac{\sqrt{a^3}}{a^3(1+x)^3}\times 10^{-5.5}\:\mathrm{Hz},
\end{equation}
which, for $a=10$ nm, is equal to $g(r)\approx 3$ MHz at a distance of 10 nm ($x=1$) from the surface of the magnet. A dispersive interaction follows in the usual dispersive regime where the NV-center spin and magnon mode are far detuned relative to the strength of their interaction. The potential coupling strength found here for $a=10$ nm is significantly larger than that found in Ref.~\cite{romling2023resolving} for an NV-center spin coupled via magnetic dipole coupling to the magnetic moments of interfacial lattice sites.

\end{document}